\documentclass[12pt]{article}

\usepackage{graphicx}
\usepackage{times}

\title{Ideal Isotropic Auxetic Networks From Random Networks}

\author
{Daniel R Reid\textit{$^{a\ddag}$}, Nidhi Pashine\textit{$^{b\ddag}$}, Alec S Bowen\textit{$^{a}$}, Sidney R Nagel \textit{$^{b}$},\\Juan J de Pablo\textit{$^{a}$}$^{\ast}$\\
\\
\normalsize{$^{a}$Institute for Molecular Engineering,}
\\
\normalsize{University of Chicago, Chicago IL, 60637}\\
\normalsize{$^{b}$Department of Physics and the James Franck and Enrico Fermi Institutes,}
\\
\normalsize{University of Chicago, Chicago IL, 60637}\\
\\
\normalsize{$^{\ddag}$ Equal contribution.}
\\
\normalsize{$^\ast$Correspondence and requests for materials should be addressed to: depablo@uchicago.edu.}
}

\date{}

\begin{document}


\baselineskip14pt


\maketitle

\begin{abstract}
Auxetic materials are characterized by a negative Poisson's ratio, $\nu$. As the Poisson's ratio becomes negative and approaches the lower isotropic mechanical limit of $\nu = -1$, materials show enhanced resistance to impact and shear, making them suitable for applications ranging from robotics to impact mitigation. Past experimental efforts aimed at reaching the $\nu=-1$ limit have resulted in highly anisotropic materials, which show a negative Poisson's ratio only when subjected to deformations along specific directions.  Isotropic designs have only attained moderately auxetic behavior, or have led to structures that cannot be manufactured in 3D. Here, we present a design strategy to create isotropic structures from disordered networks that leads to Poisson's ratios as low as $\nu=-0.98$. The materials conceived through this approach are successfully fabricated in the laboratory and behave as predicted. The Poisson's ratio $\nu$ is found to depend on network structure and bond strengths; this sheds light on the structural motifs that lead to auxetic behavior. The ideas introduced here can be generalized to 3D, a wide range of materials, and a spectrum of length scales, thereby providing a general platform that could impact technology.

\end{abstract}

\section*{Introduction}

Precise manipulation of the mechanical properties of solids is critical to the design of a wide spectrum of materials.  Auxetic materials, which have a negative Poisson's ratio, $\nu< 0$, represent a promising yet under-exploited class of systems for potential applications in areas such as impact mitigation\cite{alderson1994auxetic, sanami2014auxetic}, filtration \cite{alderson2000auxetic, alderson2001auxetic}, fabric design \cite{alderson2012auxetic, hu2011development}. More generally, technologies in which materials must maintain shape under deformation, including aerospace technologies\cite{liu2006literature, alderson2007auxetic}, provide fertile grounds for the use of auxetic systems.

A variety of auxetic materials have been proposed in recent years; examples range from metamaterials\cite{schwerdtfeger2011design, shan2015design, prall1997properties, larsen1997design, alderson2007auxetic} and foams prepared by special processing techniques\cite{lakes1987foam, friis1988negative, smith2000novel, alderson1994auxetic} to composites\cite{evans1991design}.  The vast majority of these materials are anisotropic, meaning that the Poisson's ratio, $\nu$, depends on the direction of applied strain; this restriction is not desirable for applications.  Of the few materials that are isotropic, all except specially prepared foams, which can reach $\nu=-0.82$\cite{smith2000novel}, are inherently two-dimensional or exceptionally complex\cite{robert1985isotropic}, thereby rendering them difficult, if not impossible to manufacture.  To be widely useful, auxetics should be readily fabricated in three dimensions, and show isotropic Poisson's ratios approaching $\nu=-1$\cite{alderson1994auxetic}. Designing such systems continues to represent a grand challenge for materials research.

\begin{figure*}
\includegraphics[width=\textwidth]{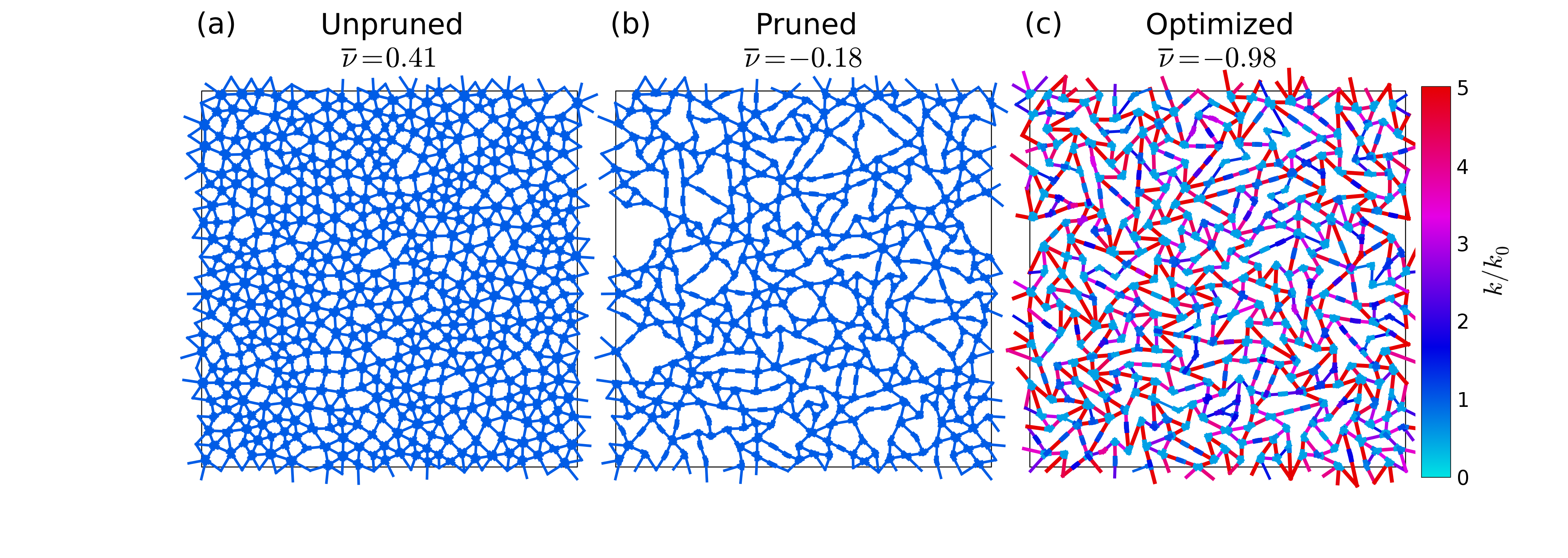}
\caption{Networks at three steps in the materials design process.  In Panel (a), an unpruned network with $Z=5.2$ and $\nu=+0.41$.  In Panel (b), the same network after low-$\Delta G_{i}$ bonds have been pruned, with $\nu=-0.18$. Panel (c) shows the same network after optimization, with $\nu=-0.98$.
The color of a bond near the node represents that bond's relative angular strength, $k_{i}^{ang}/k_{i, 0}^{ang}$, while the color along the remaining length of the bond represents its relative compressive strength, $k_{i}^{comp}/k_{i, 0}^{comp}$.  As can be appreciated, most values of $k_{i}^{comp}$ are increased while those of $k_{i}^{ang}$ are decreased.}
\label{fig:opt_diagram}
\end{figure*}

Disordered networks derived from jammed packings present an appealing means of designing auxetic materials that satisfy these elusive criteria. Such networks can be viewed as a collection of nodes connected by bonds. Previous work has shown that disordered spring networks, similar to that shown in Fig.~\ref{fig:opt_diagram}a, can be tuned to exhibit an auxetic response through selective pruning of bonds \cite{goodrich2015principle,hexner2018role,reid2018auxetic}.
In two dimensions, $\nu$ is a monotonic function of the ratio of the shear, $G$, to bulk, $B$, modulus:   $\nu=\frac{1-G/B}{1+G/B}$, where $\nu$ and $G$ are both isotropic.

In the simplest crystalline solids, the change in $G$ or $B$ upon removal of a single bond $i$ (termed $\Delta G_{i}$ or $\Delta B_{i}$ respectively) is identical for every bond.  As a consequence, removing any bond does not change $G/B$ significantly.
Disordered networks differ in two essential ways: First, the distributions of $\Delta G_{i}$ and $\Delta B_{i}$ span many orders of magnitude.  Second, these quantities for any specific bond are, to a large extent, uncorrelated with one another \cite{goodrich2015principle,hexner2018role,hexner2017linking,reid2018auxetic}.   Therefore, by iteratively removing the greatest $\Delta B_{i}$ or smallest $\Delta G_{i}$ bond, one can drive $\nu$ to negative values.  In particular, it has been shown that iterative pruning of the smallest $\Delta G_{i}$  from  disordered spring networks leads to materials with $\nu<-0.8$ in both two and three dimensions \cite{hexner2018role}. The networks considered in that work had only harmonic spring interactions between nodes. Experimental realizations, however, typically have angle-bending forces as well. When networks with such angle-bending forces were pruned, the Poisson's ratio reached only $\nu=-0.2$ for isotropic systems \cite{reid2018auxetic}.  Lower values of $\nu$ could only be obtained if the material was allowed to become anisotropic.

The fundamental question that arises then is whether it is possible to create isotropic, perfectly auxetic networks with $\nu$ approaching $-1$ by pruning random networks. In what follows, we show that by removing some of the constraints that were inherent to past design strategies, we show here that one can indeed create isotropic networks with $\nu\approx-0.98$. More specifically, we use materials-optimization techniques to augment the previous bond-removal strategies~\cite{reid2018auxetic,hexner2018role} in two distinct ways: first, we modify node positions, which changes the network geometry and, second, we modify individual bond strengths. 

Following this protocol, auxetic networks are created in the laboratory and have the predicted response. These results provide a framework for the production of highly-auxetic isotropic materials that, importantly, can be readily implemented in three dimensions.

During optimization, the network geometry changes to create concave polygons.  We show that these concave polygons correlate well with $\nu$. Auxetic networks also show a high degree of mechanical heterogeneity. Compressive strains on these networks change the area of individual polygons that constitute the network. As $\nu$ becomes more negative, changes in the area of polygons become highly disperse.  In addition, as $\nu \rightarrow -1$, networks show regions of negative moduli, serving to highlight the complex mechanics of these materials.

\section*{Results}

\subsection*{Disordered networks}
Disordered networks, the starting platform for our materials optimization process, are created from jammed packings as described in detail in previous work \cite{goodrich2015principle,hexner2018role,hexner2017linking,reid2018auxetic}.  Briefly, polydisperse hyperspheres are randomly positioned in any given simulation domain.  Particles interact via a soft-sphere potential:
\begin{equation}
V(r_{ij}) = \frac{V_0}{2} (1-\frac{r_{ij}}{\sigma_{ij}})^{2} \Theta\left(1-\frac{r_{ij}}{\sigma_{ij}}\right).
\label{eqn:softSphere}
\end{equation}
where $r_{ij}$ is the distance between particles $i$ and $j$, $\sigma_{ij}$ sets the length scale of interaction, $\Theta$ is the Heaviside step function and $V_0$ sets the energy scale. The density is chosen to control the average coordination number, $Z$, which has an important effect on network properties \cite{goodrich2015principle,reid2018auxetic}.
The system is allowed to relax to its closest energy minimum and bonds (unstretched springs) are attached between all pairs of particles for which $r<\sigma_{a}+\sigma_{b}$, where $\sigma_{a}$and $\sigma_{b}$ are the radii of particles $a$ and $b$.  The soft-sphere potential is then removed to form the network, consisting of nodes and springs, with no stresses in the system.

\subsection*{Network model}
Having formed a network, the constituent bonds are described by two potentials - harmonic compression along the length of the bond, and harmonic bending about the node.
\begin{equation}
V_{comp}(r_{ab}) = \frac{k^{comp}_{i}}{2r_{ab}^{0}} (r_{ab}-r_{ab}^{0})^{2} \;.
\label{eqn:compression}
\end{equation}

\begin{equation}
V_{bend}(\theta_{ab\vec s_b}) = \frac{k^{ang}_{i}}{2} (\theta _{ab\vec s_b}-\theta_{ab\vec s_b}^{0}) ^{2}
\label{eqn:anglepotential}
\end{equation}

The coefficient associated with bond compression for the $i$th bond is denoted $k_{i}^{comp}$ while that for bond bending is denoted $k_{i}^{ang}$.  The length of bond $i$, that connects nodes $a$ and $b$, is denoted $r_{ab}$, and $r_{ab}^{0}$ represents the unstretched length of that bond.  The bond bending potentials couple to a director on each node, $b$, which is denoted by $\vec s_b$.
These simple potentials have been shown to capture experimental network behavior accurately both in the linear regime and also at high compressive strains \cite{reid2018auxetic}.

\subsection*{Network formation and pruning}
\begin{figure}
\begin{center}
\includegraphics[width=9cm]{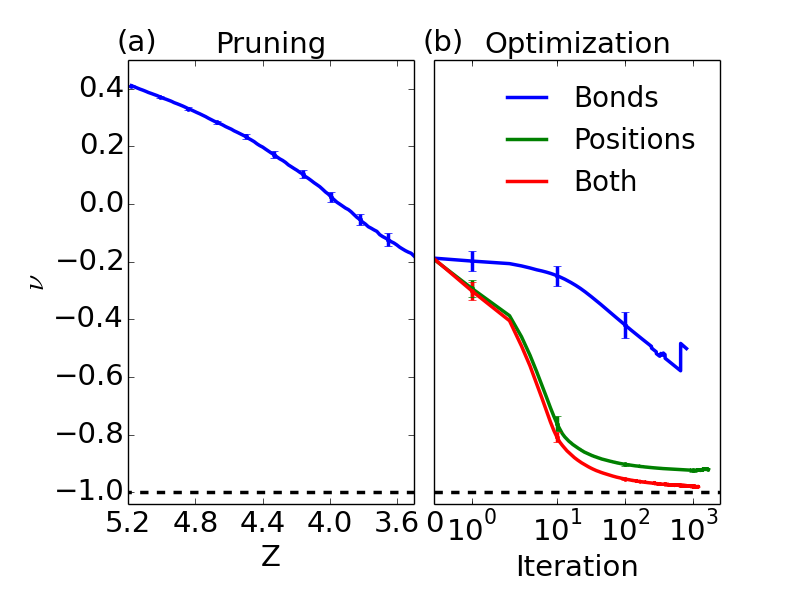}
\caption{Poisson's ratio, $\nu$, over the pruning and optimization process.  Panel a shows $\nu$ as the network is pruned from $Z=5.2$ to $Z=3.5$.  Pruned networks are then optimized as shown in Panel b.  The three data sets show $\nu$ as (i) only bond strengths are optimized, (ii) only node positions are optimized, and (iii) both bond strengths and node positions are optimized.}
\label{fig:poissons_with_opt}
\end{center}
\end{figure}

After network formation, two distinct steps are followed to create highly-auxetic isotropic materials:  (i) pruning and (ii) optimization.  An example network after each of these steps is depicted in the three panels of Fig.~\ref{fig:opt_diagram}.
Unpruned periodic networks, as exemplified in Fig.~\ref{fig:opt_diagram}a, are initialized with $Z=5.2$ and show $\overline{\nu}=+0.41$. An initial value of $Z=5.2$ has been shown to produce maximally auxetic pruned networks \cite{reid2018auxetic}.  In two dimensions, there are two independent shear moduli, $G^{p}$ and $G^{s}$: pure and simple shear.  In previous work on networks with angle-bending forces, bonds having the lowest value of $\Delta G^{p}_{i}$ were pruned, while ignoring contributions to $G^{s}$ \cite{reid2018auxetic}. Because the pruning criteria focused only on one modulus, a low value, $\nu=-0.9$, could be achieved anisotropically (auxetic with respect to strain along the principle axes, but not with respect to strain along the diagonals).  In order to form {\em isotropic} networks, the pruning is carried out as in ref.~\cite{hexner2018role} by pruning bonds having the lowest contribution to the average shear modulus $(\Delta G^{p}_{i} + \Delta G^{s}_{i})$.
The process is repeated until $Z$ is reduced to $3.5$ and $\nu=-0.18$, as shown in Fig.~\ref{fig:poissons_with_opt}a.  An example of a pruned network is shown in Fig.~\ref{fig:opt_diagram}b.  Further pruning does not significantly reduce $\nu$.

\subsection*{Network optimization}
Once networks are pruned, an optimization process is carried out to reduce $\nu$ to a value close to the isotropic mechanical limit $\nu=-1$.  A gradient descent optimization technique is used, which optimizes the compression and angle coefficients of each bond as well as the position of each node.  At each step, $t$, the spring constants are updated according to:
\begin{equation}
k_{i}(t+1) = k_{i}(t)- \Delta_{k}  \frac{\partial \nu}{ \partial k_{i} }
\label{eqn:deltaKEqn}
\end{equation}
where $k_{i}$ is the compression or angle coefficient for the $i^{th}$ bond.
Values of $k_{i}$ are constrained to lie between 0.5 and 5.0 times their initial value, to ensure that their values are not too disperse for experimental realizations. The parameter $\Delta_{k}$ is the step size of the optimization, and is set so that the maximum change of any coefficient is 10\% of the largest value of $k_{i}$ at each step in optimization.

Node positions are also optimized to minimize $\nu$:
\begin{equation}
r_{a}(t+1) = r_{a}(t) - \Delta_{r} \frac{\partial E_{a}}{\partial r_{a}}
\label{eqn:deltaREqn}
\end{equation}
\begin{equation}
E_{a} = \nu + \sum_{i \in bonds} \epsilon \Big(\frac{\sigma}{|r_{a}|}\Big)^{6}.
\label{eqn:Eeqn}
\end{equation}
The second term in Eq. \ref{eqn:Eeqn} is a repulsive term that ensures that nodes do not cross bonds.  The quantities $\sigma$ and $\epsilon$ set length and energy scales of the repulsion, and are chosen to be 0.3 length units and 0.05 respectively. As with bond coefficients, $\Delta_{r}$ sets that maximum step size which is chosen such that the maximum node displacement after every iteration is $0.1$, while initial bond lengths range from $1.2$ to $1.6$.

\subsection*{Optimization results}
Figure~\ref{fig:poissons_with_opt}b shows $\nu$ during three optimization procedures: (i) bond coefficients (Eq.~\ref{eqn:deltaKEqn}), (ii) node positions (Eq.~\ref{eqn:deltaREqn}), and (iii) both bond coefficients and node positions simultaneously.  All optimization algorithms yield isotropically auxetic materials.  Optimizing only bond parameters yields materials with $\overline{\nu}=-0.50$, and optimizing only node positions yields $\overline{\nu}=-0.91$.  Optimizing both quantities simultaneously yields $\overline{\nu}=-0.98$.  An example of a network in which both node positions and bond strengths have been optimized is shown in Fig. \ref{fig:opt_diagram}c. These results demonstrate that node positions play a dominant role in controlling $\nu$, a feature that bodes well for creation of experimental realizations of the networks, where node positions can be more precisely controlled than bond strengths.

\begin{figure}
\begin{center}
\includegraphics[width=11cm]{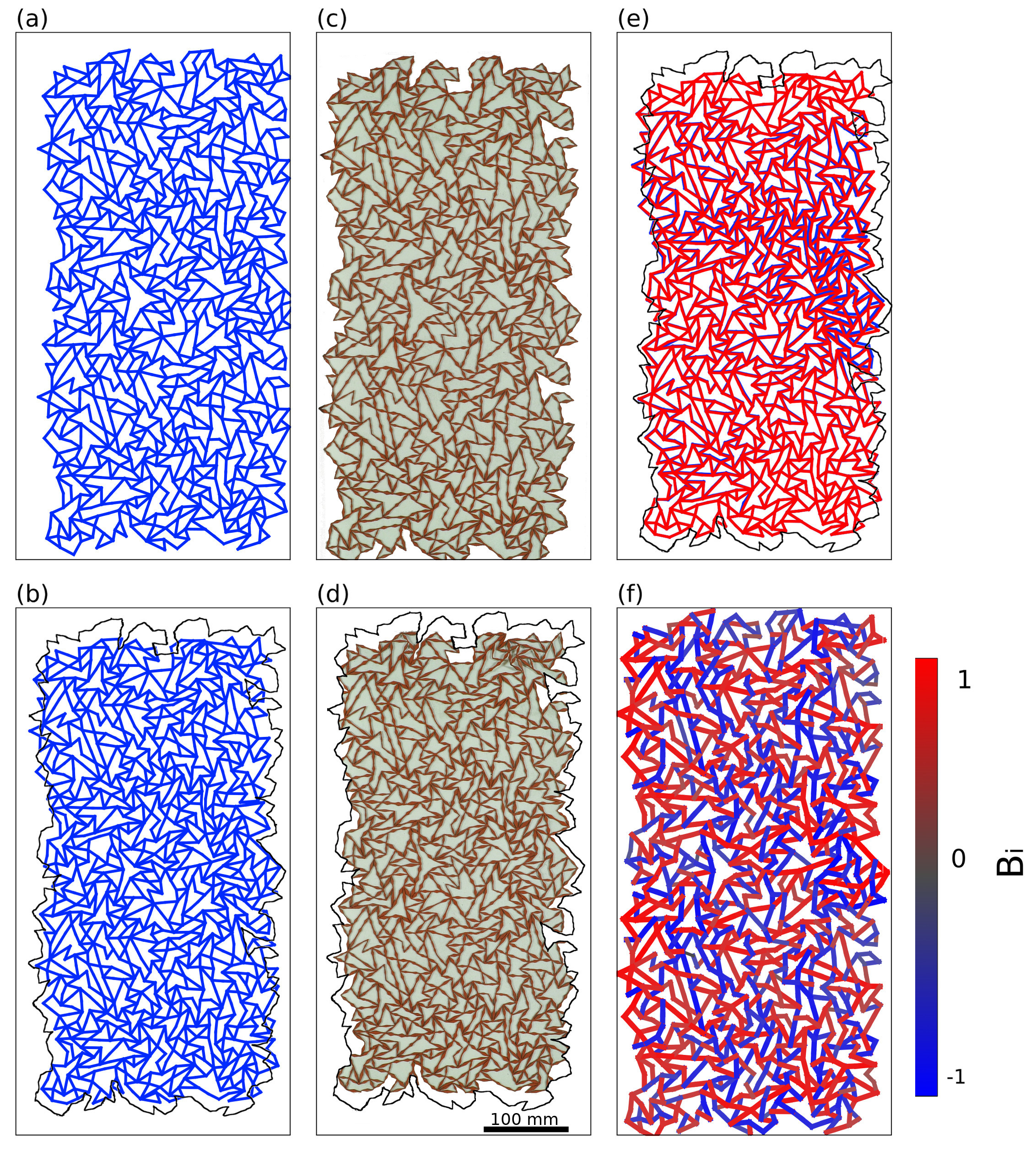}
\caption{Isotropic auxetic networks in simulation and experiment.  (a) Uncompressed simulation of auxetic network predicted to show $\nu = -0.98$.  (b) Simulated network uniaxially strained to $\epsilon_{y}=-3.6\%$ with outline of uncompressed network shown in black.  (c) Uncompressed experimental realization of identical network.  (d) Compressed experimental network at $\epsilon_{y}=-3.6\%$ with outline of uncompressed network shown in black.  (e) Comparison of experimental (red) and simulated (blue) networks at $\epsilon_{y}=-3.6\%$. The network in red is reconstructed from experimental data. (f) Bond-level contributions to $B$ in network under uniform compression, as determined by computing the virial coefficient contributions of each bond.  Contributions are normalized by the largest value of $B_{i}$.}
\label{fig:exp_network}
\end{center}
\end{figure}
\subsection*{Experimental realizations}

Experimental realizations of simulated networks are created by laser-cutting the configurations from sheets of silicone rubber.  A representative network with a simulated value of $\nu \approx -0.98$ is shown in Fig.~\ref{fig:exp_network}a.  Upon uniaxial compression in the vertical dimension ($y$), the network behaves auxetically, as shown in Fig.~\ref{fig:exp_network}b.

The experimental network, the counterpart to the computer-generated one shown in Fig.~\ref{fig:exp_network}a is shown in Fig.~\ref{fig:exp_network}c.  In order to create the  experimental network, the thicknesses and shapes of all the bonds must be carefully controlled since they are the experimental analogues of $k_{i}^{comp}$ and $k_{i}^{ang}$. In experiment, the bond compression strength, $k_{i}^{comp}$, is controlled  by the thickness of the bond along its length.  The angular strength, $k_{i}^{ang}$, on the other, hand is controlled by the thickness of the bond near its end close to a node~\cite{Rocks2017designing}.
The exact protocol adopted here for setting thicknesses and taper of bonds is described in Methods and Materials.

When the experimental network is compressed along the vertical dimension ($y$), it is auxetic as shown in Fig.~\ref{fig:exp_network}d. In Fig.~\ref{fig:exp_network}e, the compressed experimental (red) and simulated (blue) networks are superimposed. They overlap extremely well over most of the area. (The solid black outline shows the original outline of the uncompressed network.)

Recent work on atomic glasses has revealed that 
regions of negative moduli exist that influence a material's mechanical response and relaxation \cite{yoshimoto2004mechanical}. Regions of negative moduli also appear in our disordered networks.  These local moduli are defined as the virial coefficients associated with individual particles or bonds.  A bond that has a negative contribution to $B$ is under tension while the material is under compression. 
Figure \ref{fig:exp_network}f shows the per-bond contribution to $B$ for a network with $\nu=-0.8$.  Interestingly, 34\% of bonds in optimized networks contribute negatively to the bulk modulus, compared to 24\% and 2\% in pruned and unpruned networks, respectively.

To validate that the networks are isotropic, two realizations of a 500 node auxetic network were created in laboratory experiments.  The two networks were cut from the same periodic unit cell of the simulation, one as a square that lies flat along the $x$ axis, and the other lying at $45^{\circ}$ to the $x$ axis (the $xy$ axis).  In the experiment, these two networks can be compressed along the $x$ and $xy$ axes, respectively.

\begin{figure}
\begin{center}
\includegraphics[width=10cm]{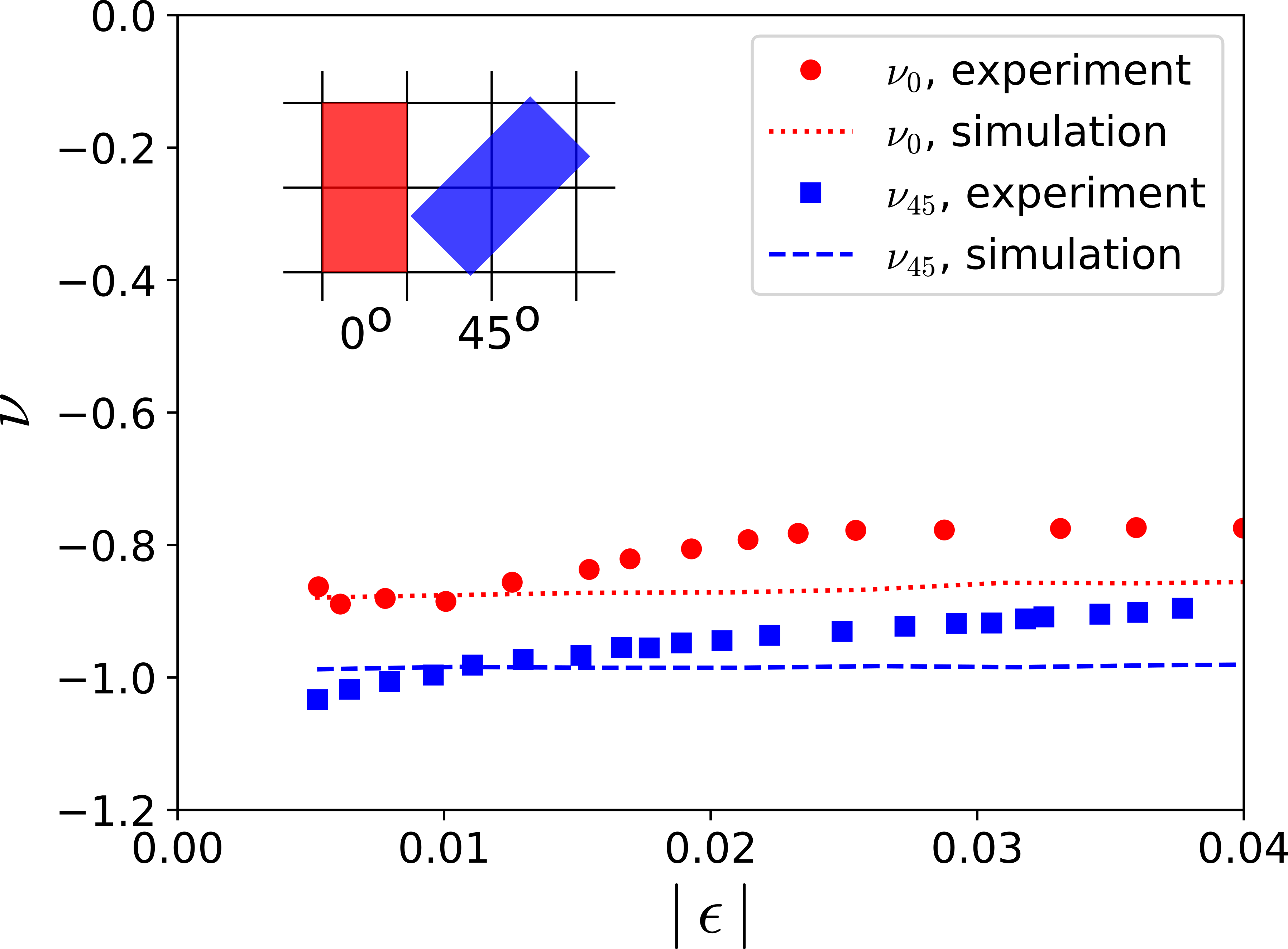}
\caption{Response of networks in simulations and experiments. Starting from the same unit cell, two different $2\times1$ tilings were made, one rotated by $45^\circ$ with respect to the other as shown in the inset. Abscissa shows the compressive strain applied along the principal axis of the two networks to determine $\nu_{0}$ (red circles, dotted line) and $\nu_{45}$ (blue squares, dashed line) respectively.  The solid points are from experiment and the lines are from simulations. }
\label{fig:exp_sim_compare}
\end{center}
\end{figure}

Deforming the first network along the $x$ (or $y$) axis probes $\nu$, as it relates to $G^{p}/B$, which we denote $\nu_{0}$.
Deforming the second network along the $xy$ (or $yx $) axis probes $\nu$ as it relates to $G^{s}/B$, which we denote $\nu_{45}$.  A material that shows the same value for $\nu_{0}$ and $\nu_{45}$ is perfectly isotropic.  As shown in Fig.~\ref{fig:exp_sim_compare}, While the networks produced here are not perfectly isotropic, they have the same value of Poisson's ratio within $\pm$ 0.1.  Moreover, they show excellent agreement between simulation and experiment: $\nu_{0}=-0.88, -0.87$ and $\nu_{45}=-0.98, -1.03$, for simulation and experiment, respectively.  The values of $\nu$ are nearly constant up to strains of $-0.04$.

\begin{figure*}
\begin{center}
\includegraphics[width=\textwidth]{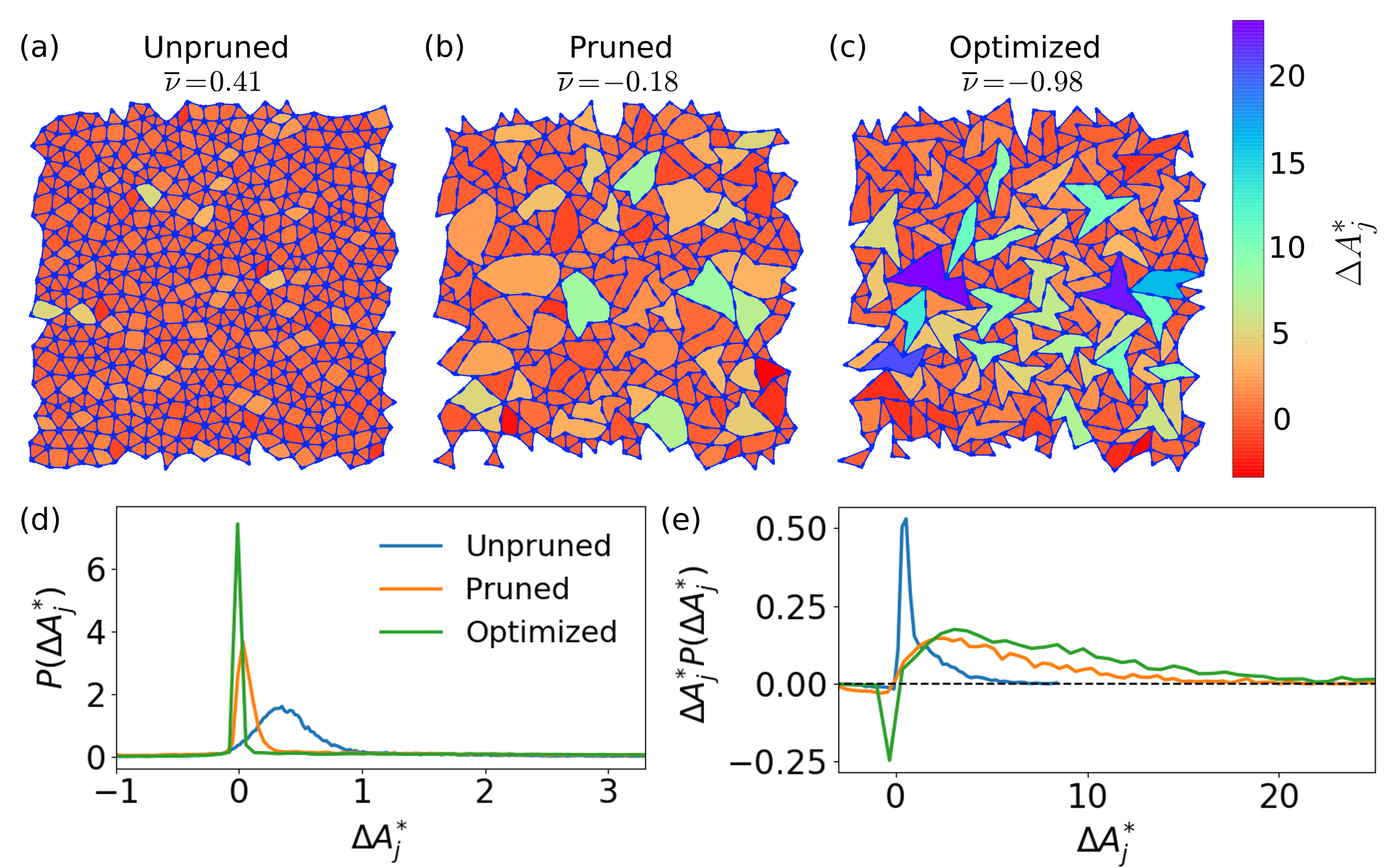}
\caption{Normalized changes in minimal polygon areas, $\Delta A_{j}^{*}$ of a single 500 node network at the three stages of preparation: (a) unpruned, (b) pruned, and (c) optimized. $\Delta A_{j}^{*}$ was measured after the networks were compressed by $\epsilon=0.01\%$.  More negative values of $\Delta A_{j}^{*}$ contribute to lowering $\nu$.  As can be observed, networks with lower values of $\nu$ are more mechanically heterogeneous. (d) Probability Distributions of minimal polygon deformations, $\Delta A_{j}^{*}$ upon uniaxial compression for unpruned, pruned, and optimized networks. As networks are pruned and then optimized, the distributions becomes increasingly sharp. (e) Probability distribution, $P(\Delta A_{j}^{*})$ weighted by $\Delta A_{j}^{*}$, demonstrates that for pruned and optimized networks, the low probability outliers of $\Delta A_{j}^{*}$ are responsible for the majority of the material deformation.}
\label{fig:d_areas}
\end{center}
\end{figure*}

\subsection*{Physical understanding of auxetic behavior}

In order to develop an understanding of the structural features that give rise to auxetic behavior, we now examine the properties of the individual polygons (also called minimal cycles) that compose the network. As shown in the Methods Section, the Poisson's ratio can be related to a change in each polygon's area due to a uniaxial compression in the $y$-direction, $\epsilon_{y}$:
\begin{equation}
\nu = \frac{1}{(\epsilon_{y}+1)} \Big(1-\frac{\sum_{j=1}^{N} \Delta A_{j}}{N \epsilon_{y} \overline{A}} \Big) = \frac{1}{(\epsilon_{y}+1)} \Big(1-\frac{1}{N}\sum_{j=1}^{N} \Delta A_{j}^{*}\Big)
\label{eqn:cycleDeform}
\end{equation}

Here, $\Delta A_{j}$ is the change in the area of the $j^{th}$ polygon, $N$ is the total number of polygons, $\overline{A}$ is the average area of all the polygons in the network and
\begin{equation}
\Delta A_{j}^{*} = \frac{\Delta A_{j}}{\epsilon_y\overline{A}}
\label{eqn:deformNormalize}
\end{equation}

Large positive values of $\Delta A_{j}^{*}$ contribute predominantly to lowering $\nu$.

Fig.~\ref{fig:d_areas}a, b, and c respectively show $\Delta A_{j}^{*}$ of each polygon in an example of an unpruned, pruned, and optimized network. In the unpruned system, the values of $\Delta A_{j}^{*}$ are relatively uniformly distributed in space and have roughly the same magnitude throughout.  The values of $\Delta A_{j}^{*}$ fluctuate much more and become more heterogeneously distributed as the networks are pruned, and even more so when optimized.  Additionally, it can be observed that all polygons with very high values of $\Delta A_{j}^{*}$ are concave, consistent with recent predictions \cite{reid2018auxetic}.  It is unclear to what extent these observations apply to other auxetic materials, however many classes of auxetics possess concave structures similar to those seen here \cite{reid2018auxetic, alderson2012auxetic, ravirala2006negative,lakes1987foam}.

Figure~\ref{fig:d_areas}d shows $P(\Delta A_j^*)$, the distributions of $\Delta A_{j}^{*}$, taken from $30$ independent networks at the three stages in the formation process.  As networks become more auxetic, a decreasing number of polygons become responsible for the auxetic behavior; most polygons show relatively little change in area when compressed.

As the network is pruned and optimized, the distribution $P(\Delta A_j^*)$ gets sharper. However, the most probable value of $\Delta A_{j}^{*}$ for unpruned, pruned, and optimized networks decreases from $0.35$, to $0.03$ to $-0.01$. This is unexpected since a more positive value of $\Delta A_j^*$ contributes to lowering the Poisson's ratio.

To understand this, the relevant quantity is the probability distribution weighted by $\Delta A_j^*$, as shown in Fig.~\ref{fig:d_areas}e. Unpruned, pruned, and optimized networks respectively have average $\Delta A_{j}^{*}$ of $0.55 \pm 0.72$, $1.04\pm 2.38$, and $1.87\pm 3.75$.
A significant contribution to the auxetic response in optimized networks is from the long positive tail of $\Delta A_{j}^{*}$ which goes as high as $25$. Another feature seen in these plots is the growing number of polygons with $\Delta A_j^* < 0$. The fraction of such polygons for unpruned, pruned, and optimized networks is $0.06$, $0.16$, and $0.17$ respectively.

These results paint a picture in which materials become highly heterogeneous as they become auxetic via this pruning and optimization process.  While very positive values of $\Delta A_{i}^{*}$ contribute to lowering $\nu$, most polygons in our auxetic networks are relatively unchanged when the material is compressed. For example, in the optimized case, only $21\%$ of the polygons have $\Delta A_j^* >3$, but they contribute to $77\%$ of the area change when the network is compressed.

The auxetic behavior we have found is largely the result of a relatively small number of highly compliant polygons.

\begin{figure}
\begin{center}
\includegraphics[width=15cm]{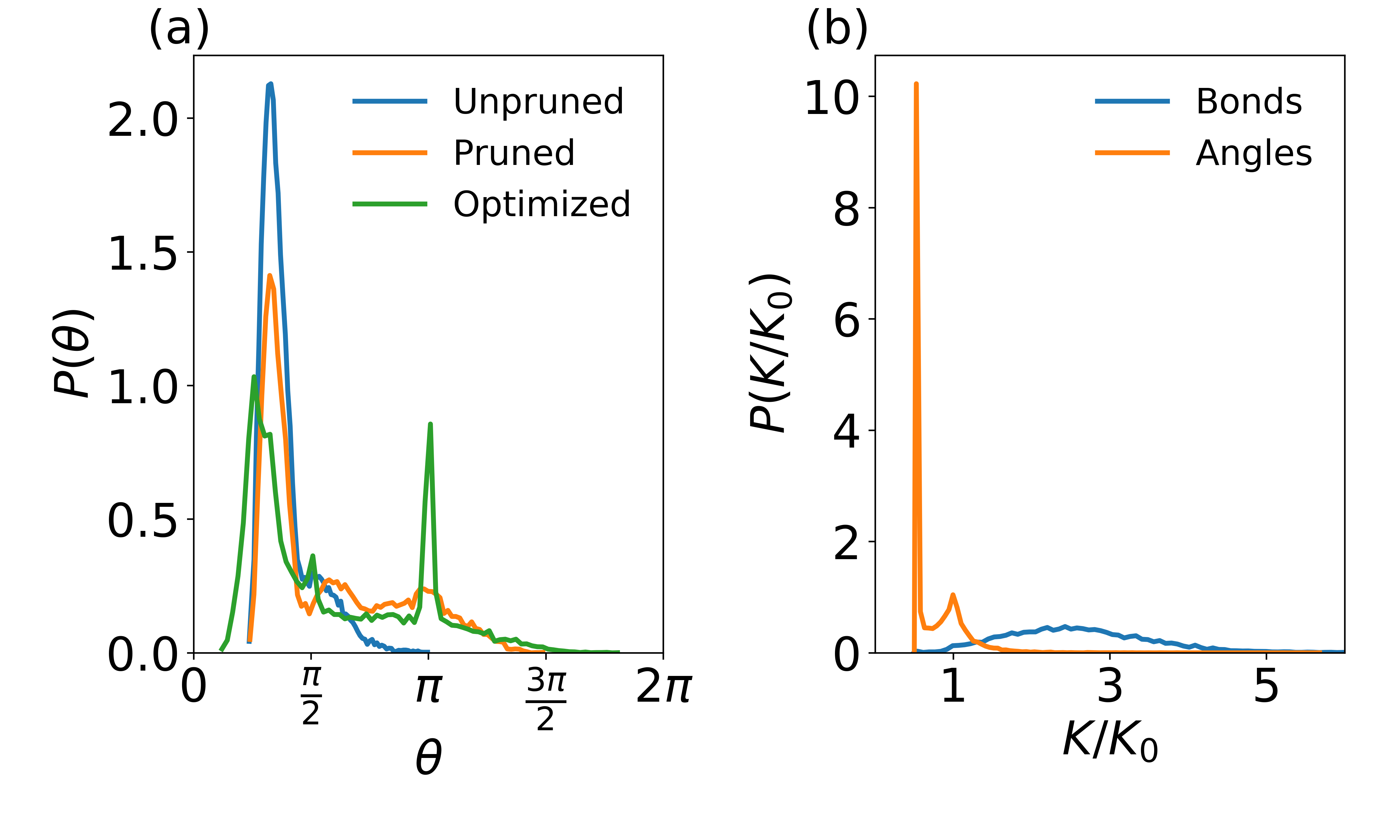}
\caption{Probability distributions of network structural features. (a) Probability distributions of angles between adjacent bonds at nodes.  (b) Probability distribution of compressive and angular coefficients for optimized networks.}

\label{fig:ml}
\end{center}
\end{figure}

As networks are pruned and optimized, we note distinct changes in their underlying structure.  Figure~\ref{fig:ml}a shows the probability distributions of angles between adjacent bonds on nodes for unpruned (Z=5.2), fully pruned (Z=3.5), and fully optimized networks.

In the unpruned network, angles are tightly distributed around $1.2$ radians, which corresponds to $2\pi/Z_{0}$ where $Z_{0}=5.2$. In pruned networks, the average angle grows to $1.8$ radians, and the distribution becomes broader.  In optimized networks, small angles become more common and a second peak appears at $\pi$.

The strength of the bonds also varies during pruning and optimization.  We measure the ratio of the spring constants on each bond before and after pruning and optimization, $k_{i}^{comp}/k_{i, 0}^{comp}$ and $k_{i}^{ang}/k_{i, 0}^{ang}$.  Figure~\ref{fig:ml}b shows probability distributions $P(k_{i}^{comp}/k_{i, 0}^{comp})$ and $P(k_{i}^{ang}/k_{i, 0}^{ang})$ for optimized networks; 82\% of the angular coefficients are weakened while a small fraction are unchanged or grow stronger.  The average angle is 0.73 times as strong as its original value.  (Note that the large peak at $0.5$ is due to the imposed minimum angle strength.)  Compressive coefficients, on the other hand, show a broad distribution with 98\% growing stronger.  The average $k_{i}^{comp}$ is 2.6 times stronger than its original value.

\section*{Discussion}
The work presented here lays out a framework for the design of tunable, highly auxetic isotropic materials with unique mechanical properties.  The structures will be auxetic if fabricated at any size, from molecular to architectural-scales.  While fabrication for large-scale devices is relatively straightforward, fabrication at smaller scales presents a greater challenge. For small scales, one option could be to use purposefully assembled DNA-functionalized nanoparticles.  Such assemblies have been shown to have a host of unusual mechanical properties \cite{mirkin1996dna,lequieu2016mechanical} and can be made amorphous, a requirement for these materials.  Glassy materials can also show a host of unusual and tunable properties depending on material processing and formation conditions \cite{antony2017influence, reid2016age, singh2013ultrastable}. With sufficiently high resolution, 3D printing may also provide a feasible path towards material fabrication \cite{sun20133d}.  Now that several possible fabrication routes have become available, the manufacture of micro-scale auxetic networks prepared by pruning should prove a fruitful avenue for materials development.

\subsection*{Methods}
\subsubsection*{Experimental methods}
Experimental systems were laser cut from rubber sheets. These are silicone rubber sheets that are $1.5mm$ thick and have a stiffness value of Shore 70A. Each bond of the network was made to have a specific thickness, depending on its value of $k_{i}^{ang}$ and $k_{i}^{comp}$. Each bond has an optimized $k_{i}^{comp}$ and two $k_{i}^{ang}$'s, one for each end of the bond.
The simulations have upper and lower bounds on the values that $k_{i}^{comp}$ and $k_{i}^{ang}$ can take. In the experimental systems, the width of each bond is kept between $1mm$ and $2mm$ near the nodes and between $2mm$ and $4mm$ at the center of the bonds. The width of bond $i$ goes as $\sqrt{k_{i}^{comp}}$ or $\sqrt{k_{i}^{ang}}$ depending on the part of the bond. In order to prevent bonds from running into each other, as we go from the middle of the bond towards a node, the central part of the bond tapers off and connects smoothly to the thin part near the node. Average bond lengths in the network are $~15mm$ and the system size is around $300mm \times 600mm$ for a $2 \times 1$ tiling of periodic networks.

\subsubsection*{Poisson's ratio as a function of minimal polygon area}
Here we relate the Poisson's ratio of a network to the area of the individual polygons. For a rectangular network, the total change in area of the entire system is a sum over the contributions from all of the polygons:
\begin{equation}
\Sigma \Delta A_j = x_f y_f - x_0 y_0
\end{equation}
where $(x_0,y_0)$ and $(x_f,y_f)$ are respectively the initial and final widths and lengths of the entire network.   For uniaxial compression along $y$ axis, the Poisson's ratio is given by:
\begin{equation}
\nu = -\frac{\epsilon_x}{\epsilon_y} = \Big(\frac{1}{\epsilon_y}\Big)\Big(1-\frac{x_f}{x_0}\Big)
\end{equation}
where $\epsilon_{y} = \frac{y_f - y_0}{y_0}$.
Substituting ${x_f}/{x_0}$ in terms of $\Sigma \Delta A_j$ gives:
\begin{equation}
\nu = \Big(\frac{1}{\epsilon_y}\Big)\Big(1-\frac{\Sigma \Delta A_j}{x_0 y_f} - \frac{y_0}{y_f}\Big) = \frac{1}{(\epsilon_{y}+1)} \Big(1-\frac{\sum_{j=1}^{N} \Delta A_{j}}{N \epsilon_{y} \overline{A}} \Big).
\end{equation}
This is the same as Eq.~\ref{eqn:cycleDeform}.

\section*{Acknowledgments}

We wish to thank Irmgard Bischofberger, Carl Goodrich, Daniel Hexner, Heinrich Jaeger and Andrea Liu for stimulating conversations.
NP was supported by NSF MRSEC DMR-1420709. SRN was supported by DOE DE-FG02-03ER46088 and the Simons Foundation for the collaboration “Cracking the Glass Problem” award $\#$348125.  The development of auxetic systems for impact mitigation applications and the corresponding materials optimization strategies presented here are supported by the Center for Hierarchical Materials Design (CHiMaD), which is supported by the National Institute of Standards and Technology, US Department of Commerce, under financial assistance award 70NANB14H012.

\bibliographystyle{unsrt}
\bibliography{main}

\end{document}